# Worst-Case and Average-Case Analysis of n-Detection Test Sets

Irith Pomeranz[1]
School of Electrical & Computer Eng.
Purdue University
W. Lafayette, IN 47907, U.S.A.

and

Sudhakar M. Reddy[2]
Electrical & Computer Eng. Dept.
University of Iowa
Iowa City, IA 52242, U.S.A.

**Abstract**

*Test sets that detect each target fault n times (n-detection test sets) are typically generated for restricted values of n due to the increase in test set size with n. We perform both a worst-case analysis and an average-case analysis to check the effect of restricting n on the unmodeled fault coverage of an (arbitrary) n-detection test set. Our analysis is independent of any particular test set or test generation approach. It is based on a specific set of target faults and a specific set of untargeted faults. It shows that, depending on the circuit, very large values of n may be needed to guarantee the detection of all the untargeted faults. We discuss the implications of these results.*

## 1. Introduction

Under an $n$-detection test set, each target fault is detected $n$ times, by $n$ different test vectors. If a fault has fewer than $n$ different test vectors that detect it, all its test vectors are included in an $n$-detection test set. The motivation for the use of $n$-detection test sets is that by increasing the number of detections of target faults, the likelihood of detecting unmodeled faults and defects is increased as well. In addition, generation of $n$-detection test sets for a specific fault model requires only minor modifications to a test generation procedure for the same fault model. This is typically simpler than writing a test generation procedure for a new fault model. Generation of $n$-detection test sets and their advantages in detecting unmodeled faults and defects were studied in [1]-[7].

Determining a value for $n$ is typically done based on tester memory and test application time constraints. Since the size of a compact $n$-detection test set increases approximately linearly with $n$, $n \leq 10$ appears to have become the accepted bound on $n$. Several questions arise when $n$ is restricted. (1) How much unmodeled fault or defect coverage is missed by restricting $n$. (2) How much higher should $n$ be in order to eliminate the fault or defect coverage loss. The latter is important in deciding whether higher values of $n$ should be considered.

A model for estimating the defective part level after the application of a given test set was given in [3] and [4].

The model is based on the numbers of times fault sites are activated and observed by the test set, which are improved by $n$-detection test generation. This model can provide answers to the questions above with respect to a given test set. However, it was not used for providing general answers that are independent of the test set or the test generation procedure used. In addition, to answer the second question it would be necessary to perform test generation for increasing values of $n$, and the bound on $n$ is unknown a-priori.

In this work we investigate a method of analysis to answer both questions in a way that is independent of any particular test set or test generation procedure. We perform the analysis by considering a specific set of target faults for $n$-detection test generation (single stuck-at faults), and a specific set of untargeted faults (four-way bridging faults [8], [9]). The four-way bridging fault model was used for developing manufacturing tests for static defects in industrial designs [9], [10]. Bridging faults were also used earlier as surrogates for unmodeled defects in estimating the defect coverage obtained by a given test set [3], [4]. The questions we answer become the following. (1) How much untargeted fault coverage (coverage of four-way bridging faults) is missed by restricting $n$. (2) How much higher should $n$ be in order to ensure that all the untargeted faults (all the four-way bridging faults) are detected.

We investigate both a worst-case analysis and an average-case analysis to answer these questions. The worst-case analysis assumes that if it is possible to generate an $n$-detection test set that will fail to detect an untargeted fault, such a test set will be generated. The average-case analysis provides the probabilities that an arbitrary $n$-detection test set will detect untargeted faults. The worst-case analysis and its results are described in Section 2. The average-case analysis and its results are described in Section 3. The results of Sections 2 and 3 are analyzed in Section 4, and additional results, using a stricter definition of an $n$-detection test set, are presented to support the analysis.

## 2. Worst-case analysis

Let $F = \{f_0, f_1, \cdots, f_{k-1}\}$ be the set of faults for which $n$-detection test generation is carried out (the set of target

―――――――――
1. Research supported in part by SRC Grant No. 2004-TJ-1244.
2. Research supported in part by SRC Grant No. 2004-TJ-1243.

faults). Let $G = \{g_0, g_1, \cdots, g_{m-1}\}$ be the set of faults through which the $n$-detection test set is evaluated (the set of untargeted faults). Our goal in this section is to obtain (1) the number of faults in $G$ that are *guaranteed* to be detected by an $n$-detection test set for $F$; and (2) the minimum value of $n$ required to *guarantee* that all the detectable faults in $G$ will be detected.

We base the analysis on the set $U$ of all the input vectors of the circuit. For every fault $h_i \in F \cup G$, $T(h_i) \subseteq U$ is the set of test vectors that detect $h_i$. The analysis based on the set of all the input vectors can be done only for circuits with small numbers of inputs. At the end of Section 4 we discuss how the proposed analysis can be used for large designs.

We first illustrate the analysis using the example circuit shown in Figure 1. We use the set of collapsed single stuck-at faults as the set $F$, and the set of four-way bridging faults as the set $G$. The fault line $l$ stuck-at $a$ is denoted by $l/a$. A four-way bridging fault is denoted by $(l_1, a_1, l_2, a_2)$. The fault is activated when $l_1 = a_1$ and $l_2 = a_2$. It then results in $l_1 = \bar{a}_1$ in the faulty circuit. We use the decimal representation of the input vectors to represent $U$. For the circuit of Figure 1 we have $U = \{0, 1, \cdots, 15\}$.

The bridging fault $g_0 = (9, 0, 10, 1)$ is detected by the set of input vectors $T(g_0) = \{6, 7\}$. In the first three columns of Table 1 we show all the single stuck-at faults in $F$ that are detected by at least one of the input vectors 6 and 7, and the set $T(f_i)$ for every such fault $f_i$. Considering $f_0 = 1/1$ with $T(f_0) = \{4, 5, 6, 7\}$, we note that it is possible to detect $f_0$ twice, using test vectors 4 and 5, without detecting $g_0$. A third detection of $f_0$ will require that either input vector 6 or input vector 7 will be included in the test set, thus guaranteeing that $g_0$ will be detected. Considering $f_1 = 2/0$ with $T(f_1) = \{6, 7, 12, 13, 14, 15\}$, it is possible to detect $f_1$ four times, using test vectors 12, 13, 14 and 15, without detecting $g_0$. A fifth detection of $f_1$ will guarantee that $g_0$ will be detected.

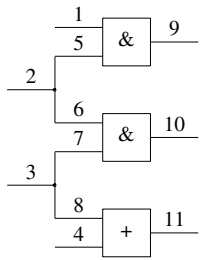

**Figure 1: Example circuit**

In general, for a fault $g_j \in G$, let $F(g_j) \subseteq F$ be the subset of faults such that $T(f_i) \cap T(g_j) \neq \phi$ for every $f_i \in F(g_j)$. Let $N(f_i)$ be the size of $T(f_i)$, and let $M(g_j, f_i)$ be the size of the intersection $T(f_i) \cap T(g_j)$. We can detect $f_i$ $N(f_i) - M(g_j, f_i)$ times without detecting $g_j$. If the number of detections of $f_i$ is increased to $N(f_i) - M(g_j, f_i) + 1$, $g_j$ is guaranteed to be detected. We denote $N(f_i) - M(g_j, f_i) + 1$ by $n_{\min}(g_j, f_i)$. This is the minimum value of $n$ for which detection of $f_i$ guarantees detection of $g_j$. Considering all the faults in $F(g_j)$, we define $n_{\min}(g_j) = \min\{n_{\min}(g_j, f_i) : f_i \in F(g_j)\}$. This is the minimum number of detections $n$ that guarantees detection of $g_j$ by an $n$-detection test set for $F$.

**Table 1: Faults with test vectors that overlap with $T(g_0) = \{6, 7\}$**

| $i$ | $f_i$ | $T(f_i)$ | $n_{\min}(g_0, f_i)$ |
|---|---|---|---|
| 0 | 1/1 | 4 5 6 7 | 3 |
| 1 | 2/0 | 6 7 12 13 14 15 | 5 |
| 3 | 3/0 | 2 6 7 10 14 15 | 5 |
| 9 | 8/0 | 2 6 10 14 | 4 |
| 11 | 9/1 | 0 1 2 3 4 5 6 7 8 9 10 11 | 11 |
| 12 | 10/0 | 6 7 14 15 | 3 |
| 14 | 11/0 | 1 2 3 5 6 7 9 10 11 13 14 15 | 11 |

In Table 1 we show the values of $n_{\min}(g_0, f_i)$ considering all the faults in $F(g_0)$. Based on the information given in Table 1, we have $n_{\min}(g_0) = 3$. Thus, in order to guarantee that an arbitrary $n$-detection test set will detect $g_0$, $n$ must be larger than or equal to three.

Results of the computation of $n_{\min}(g_j)$ for the combinational logic of MCNC finite-state machine benchmarks are shown in Tables 2 and 3 (at the end of the paper). In Table 2 we are interested in faults out of $G$ that are guaranteed to be detected by any $n$-detection test set for $n \leq n_{\max}$. Here, $n_{\max}$ is a constant for which it is practical to generate an $n$-detection test set ($n_{\max} = 10$ in our experiments). In Table 3 we are interested in faults out of $G$ that require large values of $n$ to guarantee that they will be detected by an $n$-detection test set. We report in Table 3 on faults that are guaranteed to be detected by an $n$-detection test set only if $n > n_{\max}$. We only include in Table 3 circuits for which such faults exist.

Table 2 is organized as follows. After the circuit name we show the number of four-way bridging faults considered (detectable non-feedback four-way bridging faults between outputs of multi-input gates). Under column $n_{\min}(g_j) \leq n_{min0}$ we show the percentage of faults $g_j \in G$ such that $n_{\min}(g_j) \leq n_{min0}$. This is the percentage of faults for which an $n$-detection test set that guarantees their detection requires $n \leq n_{min0}$. If $n = n_{min0}$ results in 100% coverage of four-way bridging faults, we do not report on higher values of $n$. For example, *bbara* has 858 faults in $G$. Of these faults, 80.42% require $n_{\min}(g_j) = 1$ to guarantee their detection, i.e., they will be detected by any 1-detection test set; 84.85% of the faults require $n_{\min}(g_j) \leq 2$ to guarantee their detection, i.e., they will be detected by any 2-detection test set; 97.55% of the faults are guaranteed to be detected by any 10-detection test set.

Table 3 is organized as follows. After the circuit name we show the number of four-way bridging faults considered. Under column $n_{\min}(g_j) \geq n_{min0}$ we show the number (and the percentage in parentheses) of faults $g_j \in G$ such that $n_{\min}(g_j) \geq n_{min0}$. This is the number (or percentage) of faults for which an $n$-detection test set that guarantees their detection must be computed for $n \geq n_{min0}$. For example, considering *bbara*, none of the faults in $G$ requires $n_{\min}(g_j) \geq 100$ to guarantee its detection. Three faults require $n_{\min}(g_j) \geq 20$, and 21 faults require $n_{\min}(g_j) \geq 11$ (which is 2.45% of the faults in $G$).

From Tables 2 and 3 it can be seen that large percentages of the faults in $G$ will be detected by any $n$-detection test set for $n = 1$, and very large percentages will be detected by any $n$-detection test set for $n = 10$. Nevertheless, there are non-trivial numbers of faults that are not guaranteed to be detected by a 10-detection test set. For the last seven circuits in Table 3, even $n = 100$ is not sufficient to guarantee the detection of all the faults in $G$. We show the distribution of $n_{\min}(g_j)$ for *dvram* in Figure 2, considering faults for which $n_{\min}(g_j) \geq 100$.

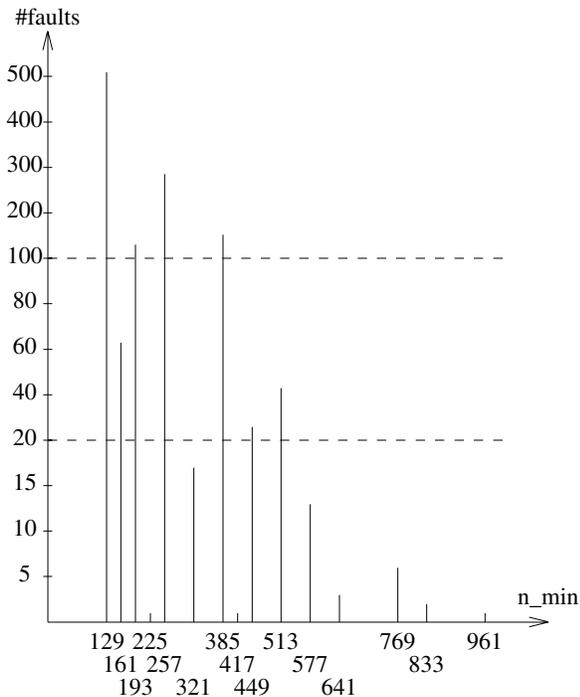

**Figure 2: Distribution of $n_{\min}(g_j)$ for *dvram***

## 3. Average-case analysis

The analysis in Section 2 is a worst-case analysis since it assumes that as long as $n$ is low enough to allow a fault $g_j \in G$ to escape detection, a test for $g_j$ will not be included in an $n$-detection test set. In practice, this worst case may not happen. Our goal in this section is to obtain the probabilities of detecting the faults in $G$ by an arbitrary $n$-detection test set, for different values of $n$.

We construct several $n$-detection test sets randomly for $n = 1, 2, \cdots, n_{\max}$ using Procedure 1 given below. Procedure 1 constructs $K$ test sets $T_0, T_1, \cdots, T_{K-1}$ for every value of $n$. Initially, $T_k = \phi$ for $0 \leq k < K$. Procedure 1 performs $n_{\max}$ iterations where it considers every test set. At the end of iteration $n$, $T_k$ is an $n$-detection test set, for $0 \leq k < K$. In iteration $n$, Procedure 1 considers every fault $f_i \in F$. For $0 \leq k < K$, if $f_i$ is detected by $T_k$ fewer than $n$ times, and $T(f_i)$ contains tests that are not included in $T_k$ (i.e., $T(f_i) - T_k \neq \phi$), Procedure 1 selects a test randomly out of $T(f_i) - T_k$, and adds it to $T_k$.

**Procedure 1:** Constructing $n$-detection test sets
(1) For $k = 0, 1, \cdots, K-1$, set $T_k = \phi$. Set $n = 1$.
(2) For every $f_i \in F$:
   For every test set $T_k$, $k = 0, 1, \cdots, K-1$:
   (a) Find the number of times $f_i$ is detected by $T_k$. Let this number be $n_{i,k}$.
   (b) If $n_{i,k} < n$ and $T(f_i) - T_k \neq \phi$:
       Select a test $t \in T(f_i) - T_k$ randomly.
       Add $t$ to $T_k$.
(3) Set $n = n+1$. If $n \leq n_{\max}$, go to Step 2.

Using the test sets $T_0, \cdots, T_{K-1}$ produced by Procedure 1 for every value of $n$, we estimate the probability that an arbitrary $n$-detection test set will detect a fault $g_j \in G$ as follows. For every $n$-detection test set $T_k$ such that $g_j$ is detected by $T_k$ (i.e., $T(g_j) \cap T_k \neq \phi$), we increment a count $d(n, g_j)$ by one. We define the probability of detecting $g_j$ by an arbitrary $n$-detection test set as $p(n, g_j) = d(n, g_j)/K$.

For illustration, we consider the example circuit of Figure 1. In Table 4 we show $n$-detection test sets for $n = 1$ and 2. For each value of $n$ we have $K = 10$ test sets. We consider the fault $g_6$ with $T(g_6) = \{12\}$. For this fault, $n_{\min}(g_6) = 4$. For $n = 1$, the set $T(g_6)$ has a non-empty intersection with $T_5$ and $T_7$. We obtain $d(1, g_6) = 2$ and $p(1, g_j) = 0.2$. For $n = 2$, the set $T(g_6)$ has a non-empty intersection with $T_0$, $T_4$, $T_5$ and $T_7$. We obtain $d(2, g_6) = 4$ and $p(2, g_j) = 0.4$.

**Table 4: Test sets for example circuit**

| | $T_k$ | |
|---|---|---|
| $k$ | $n=1$ | $n=2$ |
| 0 | 6 8 10 13 | 2 5 6 8 10 12 13 14 |
| 1 | 0 2 7 11 13 | 0 2 4 5 7 9 10 11 13 14 |
| 2 | 3 6 8 13 | 3 4 5 6 8 9 11 13 14 15 |
| 3 | 4 5 11 14 15 | 0 3 4 5 9 10 11 14 15 |
| 4 | 5 6 8 10 15 | 2 5 6 8 9 10 12 15 |
| 5 | 0 5 7 9 10 11 12 | 0 5 6 7 9 10 11 12 14 |
| 6 | 4 8 9 11 14 15 | 4 5 8 9 10 11 14 15 |
| 7 | 1 3 4 6 8 12 | 1 3 4 6 8 9 10 12 13 14 |
| 8 | 3 5 7 8 9 13 14 | 3 4 5 7 8 9 10 13 14 |
| 9 | 2 5 6 8 9 15 | 2 4 5 6 8 9 10 14 15 |

We computed $p(n, g_j)$ using $K = 10000$ test sets for $n = 1, 2, \cdots, n_{\max}$ where $n_{\max} = 10$. We report the results considering only faults $g_j$ that are not guaranteed to be detected by a 10-detection test set, and considering only circuits that have such faults. In addition, we only report the results for $n = 10$ (the detection probabilities for lower values of $n$ are lower). The results are shown in Table 5 (at the end of the paper). After the circuit name we repeat the number of faults that are not guaranteed to be detected by a 10-detection test set (the number of faults with $n_{\min}(g_j) \geq 11$). We then show the numbers of faults for which the probability of detection $p(10,g_j)$ is at least 1, 0.9, 0.8, $\cdots$, 0.1, 0. We do not enter a number for a given probability if all the faults have a higher probability of detection. The fault with the lowest probability for *ex2* has $p(10,g_j) = 0.052$. The two faults with the lowest probabilities for *bbsse* have $p(10,g_j) = 0.093$ and 0.091. The fault with the lowest probability for *cse* has $p(10,g_j) = 0.043$.

From Table 5 it can be seen that some of the faults in $G$, that are not guaranteed to be detected by a 10-detection test set, have very high probabilities of being detected by such a test set. However, there are also non-trivial numbers of faults with lower probabilities. These faults may be left undetected by a 10-detection test set. Increasing $n$ to detect such faults may sometimes require very high values of $n$ as can be seen from Table 3. Overall, the average-case analysis is consistent with the worst-case analysis in pointing to the existence of untargeted faults that are not likely to be detected by a 10-detection test set.

## 4. Discussion

The results of the previous sections indicate the following.

An $n$-detection test set for a reasonable value of $n$ (around $n = 10$) is guaranteed to detect a high percentage of the untargeted faults, and will detect most of the remaining untargeted faults with a high probability. However, there are also faults that are likely to escape detection by an $n$-detection test set. The values of $n$ required for such faults can be very high. Thus, increasing $n$ is not likely to be an effective solution for improving the effectiveness of an $n$-detection test set.

In some situations the small loss in untargeted fault coverage may be acceptable. When it is not, direct test generation for additional fault models can be carried out. Alternatively, methods to improve the effectiveness of an $n$-detection test set can be used. To illustrate this point we use an alternate method to generate $n$-detection test sets, based on a different definition of an $n$-detection test set. The definition we used so far is the standard one. For completeness, we repeat this definition next.

**Definition 1:** Let $T$ be a test set where no test is duplicated. A fault $f$ is detected $n$ times by $T$ if $T$ contains $n$ tests that detect $f$.

Under the alternate definition from [5], two tests that detect a target fault $f$ are required to be *sufficiently* different in order to be counted as different detections of $f$. The definition given in [5] is the following.

**Definition 2:** Let $T$ be a test set where no test is duplicated. A fault $f$ is detected $n$ times by $T$ if there exist $n$ tests $t_1,t_2,\cdots,t_n$ in $T$ such that the following condition is satisfied. For every $i$ and $j$ such that $1 < i \leq n$ and $1 \leq j < i$, let $t_{ij}$ be the test which is specified in bits where $t_i$ and $t_j$ are specified to the same value, and unspecified in other bits. The test $t_{ij}$ does not detect $f$.

Definition 2 simulates $f$ under a test $t_{ij}$ that contains the common bits of $t_i$ and $t_j$. If these bits are sufficient for detecting $f$, then the two tests $t_i$ and $t_j$ are considered similar, and are not counted as two detections of $f$. If $t_{ij}$ does not detect $f$, then $t_i$ and $t_j$ are considered sufficiently different, and they are counted as different detections of $f$.

We use Definition 2 in Procedure 1 for the construction of $n$-detection test sets as follows. When we find the number of detections $n_{i,k}$ of a fault $f_i$ under a test set $T_k$, we use Definition 2 to compute $n_{i,k}$. If $n_{i,k} < n$, we find the tests out of $T(f_i)-T_k$ that, if added to $T_k$, will be counted as different detections of $f_i$ according to Definition 2. If the number of detections of $f_i$ cannot reach $n$ according to Definition 2, we use Definition 1 instead. This is done to avoid situations where faults are detected much fewer than $n$ times.

Results of average-case analysis based on Definition 1 and based on Definition 2 are shown in Table 6. In this case we used $K = 1000$ test sets for the analysis. On the first row for every circuit we show the results obtained using Definition 1, and on the second row we show the results obtained using Definition 2. We target the same set of faults in both cases.

Table 6 demonstrates that Definition 2 can improve the quality of an $n$-detection test set by increasing the probability that an untargeted fault will be detected by the test set. For example, out of the 474 four-way bridging faults with $n_{\min}(g_j) \geq 11$ in the circuit *keyb*, 381 faults are detected with probability greater than or equal to 0.8 by a standard 10-detection test set generated using Definition 1, whereas 440 faults are detected with probability greater than or equal to 0.8 if the test set is generated using Definition 2. The probabilities of detection given in Tables 5 and 6 can be used to calculate the probability that an untargeted fault escapes detection.

As noted earlier, the analysis proposed requires information regarding which targeted and untargeted faults are detected by each input vector of the circuit, and hence can be directly used for circuits with small numbers of inputs. However, as the data in Table 6 shows, the analysis can be used to evaluate the relative effectiveness of different $n$-detection test generation methods. Additionally, one can partition a larger circuit into smaller subcircuits and apply the analysis to the subcircuits. Such analysis can be used to evaluate the effectiveness of a chosen value of $n$, and to estimate the probability of untargeted faults escaping detection.

## 5. Concluding remarks

In order to restrict the test set size, $n$-detection test sets are typically generated for restricted values of $n$. We addressed the following questions related to this restriction. (1) How much unmodeled fault or defect coverage is missed by restricting $n$. (2) How much higher should $n$ be in order to eliminate the fault or defect coverage loss. Our analysis was independent of any particular test set. We used single stuck-at faults as target faults, and four-way bridging faults as untargeted faults for the purpose of this study.

We used both a worst-case analysis and an average-case analysis to provide answers for these questions. The worst-case analysis assumed that if it is possible to generate an $n$-detection test set that will fail to detect an untargeted fault, such a test set will be generated. The average-case analysis provided the probabilities for an arbitrary $n$-detection test set to detect untargeted faults.

Table 2: Worst-case percentages of detected faults (small $n$)

| circuit | faults | $n_{\min}(g_j) \leq$ | | | | | |
|---|---|---|---|---|---|---|---|
| | | 1 | 2 | 3 | 4 | 5 | 10 |
| lion | 23 | 100.00 | | | | | |
| dk27 | 218 | 83.03 | 100.00 | | | | |
| ex5 | 1287 | 92.07 | 100.00 | | | | |
| train4 | 8 | 75.00 | 100.00 | | | | |
| bbtas | 155 | 89.68 | 94.84 | 100.00 | | | |
| dk15 | 1544 | 97.99 | 99.42 | 100.00 | | | |
| dk512 | 1127 | 92.72 | 99.91 | 100.00 | | | |
| dk14 | 3694 | 90.80 | 97.64 | 99.97 | 100.00 | | |
| dk17 | 1244 | 94.21 | 98.95 | 99.92 | 100.00 | | |
| firstex | 288 | 83.33 | 97.57 | 99.65 | 100.00 | | |
| lion9 | 182 | 79.67 | 89.56 | 96.15 | 100.00 | | |
| mc | 356 | 87.08 | 92.42 | 96.35 | 100.00 | | |
| dk16 | 40781 | 92.90 | 98.75 | 99.61 | 99.94 | 100.00 | |
| modulo12 | 448 | 63.62 | 84.82 | 93.30 | 99.11 | 100.00 | |
| s8 | 294 | 59.18 | 70.41 | 95.24 | 99.32 | 100.00 | |
| tav | 176 | 51.14 | 73.86 | 88.64 | 92.05 | 100.00 | |
| donfile | 11956 | 85.95 | 97.58 | 98.59 | 99.37 | 99.79 | 100.00 |
| ex7 | 1358 | 90.65 | 97.05 | 99.26 | 99.34 | 99.34 | 100.00 |
| train11 | 482 | 69.92 | 80.08 | 92.95 | 99.59 | 99.79 | 100.00 |
| beecount | 804 | 89.30 | 97.39 | 98.51 | 98.76 | 99.25 | 99.75 |
| ex2 | 11499 | 90.30 | 96.54 | 98.57 | 99.41 | 99.78 | 99.99 |
| ex3 | 2104 | 86.26 | 95.01 | 98.95 | 99.62 | 99.76 | 99.86 |
| ex6 | 4051 | 94.20 | 94.20 | 95.51 | 95.51 | 98.52 | 99.61 |
| mark1 | 2469 | 89.67 | 89.83 | 92.99 | 93.20 | 94.53 | 95.95 |
| bbara | 858 | 80.42 | 84.85 | 89.28 | 89.51 | 92.31 | 97.55 |
| ex4 | 2038 | 88.86 | 88.86 | 89.99 | 89.99 | 93.57 | 95.98 |
| keyb | 20894 | 88.27 | 91.17 | 93.61 | 93.99 | 95.03 | 97.73 |
| opus | 1901 | 79.22 | 83.96 | 89.90 | 92.00 | 93.42 | 97.42 |
| bbsse | 4265 | 89.14 | 89.14 | 89.17 | 89.17 | 92.19 | 95.97 |
| cse | 9110 | 93.61 | 93.61 | 95.16 | 95.16 | 98.25 | 99.13 |
| dvram | 14737 | 88.78 | 88.78 | 88.78 | 88.78 | 88.78 | 88.78 |
| fetch | 8958 | 92.10 | 92.10 | 92.10 | 92.10 | 92.10 | 92.10 |
| log | 4290 | 95.36 | 95.36 | 95.36 | 95.36 | 95.36 | 95.36 |
| rie | 24150 | 95.04 | 95.04 | 95.04 | 95.04 | 95.04 | 95.04 |
| s1a | 49524 | 84.34 | 84.34 | 84.59 | 84.59 | 85.68 | 88.02 |

Table 3: Worst-case numbers of detected faults (large $n$)

| circuit | faults | $n_{\min}(g_j) \geq$ | | | | | |
|---|---|---|---|---|---|---|---|
| | | 100 | | 20 | | 11 | |
| beecount | 804 | 0 | (0.00) | 0 | (0.00) | 2 | (0.25) |
| ex2 | 11499 | 0 | (0.00) | 0 | (0.00) | 1 | (0.01) |
| ex3 | 2104 | 0 | (0.00) | 0 | (0.00) | 3 | (0.14) |
| ex6 | 4051 | 0 | (0.00) | 0 | (0.00) | 16 | (0.39) |
| mark1 | 2469 | 0 | (0.00) | 0 | (0.00) | 100 | (4.05) |
| bbara | 858 | 0 | (0.00) | 3 | (0.35) | 21 | (2.45) |
| ex4 | 2038 | 0 | (0.00) | 19 | (0.93) | 82 | (4.02) |
| keyb | 20894 | 0 | (0.00) | 206 | (0.99) | 474 | (2.27) |
| opus | 1901 | 0 | (0.00) | 4 | (0.21) | 49 | (2.58) |
| bbsse | 4265 | 2 | (0.05) | 38 | (0.89) | 172 | (4.03) |
| cse | 9110 | 2 | (0.02) | 37 | (0.41) | 79 | (0.87) |
| dvram | 14737 | 1256 | (8.52) | 1653 | (11.22) | 1653 | (11.22) |
| fetch | 8958 | 688 | (7.68) | 708 | (7.90) | 708 | (7.90) |
| log | 4290 | 199 | (4.64) | 199 | (4.64) | 199 | (4.64) |
| rie | 24150 | 1136 | (4.70) | 1197 | (4.96) | 1197 | (4.96) |
| s1a | 49524 | 258 | (0.52) | 4260 | (8.60) | 5934 | (11.98) |

**Table 5: Average-case probabilities of detection**

| | | | | | | | $p(10,g_j) \geq$ | | | | |
|---|---|---|---|---|---|---|---|---|---|---|---|
| circuit | faults | 1 | 0.9 | 0.8 | 0.7 | 0.6 | 0.5 | 0.4 | 0.3 | 0.2 | 0.1 | 0.0 |
| beecount | 2 | 0 | 0 | 0 | 0 | 0 | 0 | 0 | 0 | 0 | 2 | |
| ex2 | 1 | 0 | 0 | 0 | 0 | 0 | 0 | 0 | 0 | 0 | 0 | 1 |
| ex3 | 3 | 0 | 0 | 0 | 0 | 0 | 0 | 0 | 0 | 0 | 2 | 3 |
| ex6 | 16 | 0 | 14 | 15 | 15 | 15 | 15 | 15 | 15 | 16 | | |
| mark1 | 100 | 42 | 86 | 93 | 95 | 98 | 98 | 98 | 100 | | | |
| bbara | 21 | 3 | 14 | 16 | 17 | 18 | 19 | 20 | 20 | 21 | | |
| ex4 | 82 | 32 | 82 | | | | | | | | | |
| keyb | 474 | 100 | 371 | 383 | 418 | 419 | 429 | 434 | 443 | 445 | 453 | 474 |
| opus | 49 | 13 | 40 | 46 | 47 | 49 | | | | | | |
| bbsse | 172 | 77 | 143 | 147 | 150 | 152 | 153 | 153 | 153 | 156 | 170 | 172 |
| cse | 79 | 39 | 76 | 77 | 77 | 77 | 77 | 77 | 77 | 78 | 78 | 79 |
| dvram | 1653 | 898 | 1498 | 1530 | 1562 | 1576 | 1610 | 1610 | 1618 | 1623 | 1637 | 1653 |
| fetch | 708 | 436 | 680 | 693 | 695 | 696 | 705 | 705 | 706 | 708 | | |
| log | 199 | 68 | 167 | 172 | 172 | 172 | 172 | 172 | 193 | 193 | 199 | |
| rie | 1197 | 512 | 1046 | 1067 | 1070 | 1070 | 1134 | 1134 | 1134 | 1179 | 1179 | 1197 |
| s1a | 5934 | 2663 | 4982 | 5258 | 5434 | 5511 | 5599 | 5658 | 5772 | 5816 | 5881 | 5934 |

**Table 6: Average-case probabilities of detection under Definitions 1 and 2**

| | | | | | | | | $p(10,g_j) \geq$ | | | | | |
|---|---|---|---|---|---|---|---|---|---|---|---|---|---|
| circuit | faults | def | 1 | 0.9 | 0.8 | 0.7 | 0.6 | 0.5 | 0.4 | 0.3 | 0.2 | 0.1 | 0.0 |
| bbara | 21 | 1 | 8 | 14 | 16 | 16 | 18 | 19 | 20 | 21 | | | |
| | | 2 | 10 | 18 | 19 | 20 | 21 | | | | | | |
| bbsse | 172 | 1 | 115 | 143 | 147 | 150 | 152 | 153 | 153 | 153 | 156 | 170 | 172 |
| | | 2 | 130 | 156 | 161 | 163 | 164 | 165 | 165 | 165 | 165 | 171 | 172 |
| beecount | 2 | 1 | 0 | 0 | 0 | 0 | 0 | 0 | 0 | 0 | 2 | | |
| | | 2 | 0 | 0 | 0 | 0 | 0 | 0 | 2 | | | | |
| cse | 79 | 1 | 50 | 76 | 77 | 77 | 77 | 77 | 77 | 77 | 78 | 78 | 79 |
| | | 2 | 66 | 78 | 78 | 78 | 78 | 78 | 78 | 78 | 79 | | |
| dvram | 1653 | 1 | 1076 | 1493 | 1532 | 1564 | 1573 | 1610 | 1611 | 1618 | 1623 | 1637 | 1653 |
| | | 2 | 1382 | 1597 | 1620 | 1625 | 1628 | 1635 | 1635 | 1643 | 1645 | 1649 | 1653 |
| ex2 | 1 | 1 | 0 | 0 | 0 | 0 | 0 | 0 | 0 | 0 | 0 | 0 | 1 |
| | | 2 | 0 | 0 | 0 | 0 | 0 | 0 | 0 | 0 | 0 | 0 | 1 |
| ex3 | 3 | 1 | 0 | 0 | 0 | 0 | 0 | 0 | 0 | 0 | 0 | 2 | 3 |
| | | 2 | 0 | 0 | 0 | 0 | 0 | 0 | 0 | 0 | 0 | 2 | 3 |
| ex4 | 82 | 1 | 40 | 82 | | | | | | | | | |
| | | 2 | 61 | 82 | | | | | | | | | |
| ex6 | 16 | 1 | 1 | 14 | 15 | 15 | 15 | 15 | 15 | 15 | 16 | | |
| | | 2 | 2 | 16 | | | | | | | | | |
| fetch | 708 | 1 | 564 | 680 | 695 | 695 | 696 | 701 | 705 | 708 | | | |
| | | 2 | 601 | 693 | 704 | 704 | 705 | 705 | 705 | 708 | | | |
| keyb | 474 | 1 | 186 | 371 | 381 | 418 | 424 | 429 | 434 | 443 | 446 | 453 | 474 |
| | | 2 | 274 | 426 | 440 | 458 | 461 | 465 | 468 | 470 | 473 | 473 | 474 |
| log | 199 | 1 | 121 | 167 | 172 | 172 | 172 | 172 | 172 | 193 | 193 | 199 | |
| | | 2 | 164 | 193 | 193 | 193 | 199 | | | | | | |
| mark1 | 100 | 1 | 50 | 87 | 93 | 95 | 98 | 98 | 98 | 100 | | | |
| | | 2 | 59 | 90 | 95 | 100 | | | | | | | |
| opus | 49 | 1 | 15 | 40 | 46 | 47 | 49 | | | | | | |
| | | 2 | 27 | 44 | 46 | 47 | 49 | | | | | | |
| rie | 1197 | 1 | 796 | 1046 | 1068 | 1070 | 1070 | 1134 | 1134 | 1134 | 1179 | 1179 | 1197 |
| | | 2 | 1071 | 1144 | 1146 | 1169 | 1177 | 1179 | 1179 | 1179 | 1179 | 1179 | 1197 |
| s1a | 5934 | 1 | 3412 | 4984 | 5263 | 5432 | 5495 | 5599 | 5659 | 5777 | 5833 | 5879 | 5934 |
| | | 2 | 4617 | 5566 | 5673 | 5756 | 5794 | 5855 | 5891 | 5912 | 5924 | 5925 | 5934 |